\documentclass[twocolumn]{aastex63}

\DeclareUnicodeCharacter{2212}{-}

\usepackage{hyperref}

\providecommand{\e}[1]{\ensuremath{\times 10^{#1}}}
\newcommand{\Msun}{\ensuremath{{M_{\odot}}}}
\newcommand{\Lsun}{\ensuremath{{L_{\odot}}}}

\defcitealias{alb14}{A14}
\defcitealias{alb16}{A16}
\defcitealias{sco16}{S16}
\defcitealias{sco17}{S17}

\graphicspath{{./}{figures/}}

\received{November 8, 2021}
\revised{December 15, 2021}
\accepted{December 27, 2021}
\submitjournal{ApJ}

\shorttitle{Molecular Gas Deficiencies in Clusters}
\shortauthors{Alberts et al.}

\begin{document}

\title{Significant Molecular Gas Deficiencies in Star-forming Cluster Galaxies at $z\sim1.4$}

\correspondingauthor{Stacey Alberts}
\email{salberts@arizona.edu}

\author[0000-0002-8909-8782]{Stacey Alberts}
\affil{Steward Observatory, 
University of Arizona, 
933 N. Cherry
Tucson, AZ 85721 USA}

\author{J\'{e}a Adams}
\affiliation{Center for Astrophysics $|$ Harvard \& Smithsonian, 60 Garden St, Cambridge, MA 02138, USA}
\affiliation{Amherst College Department of Physics and Astronomy, PO Box 5000, Amherst, MA 01002-5000, USA}

\author{Benjamin Gregg}
\affiliation{Department of Astronomy, University of Massachusetts, 710 North Pleasant Street, Amherst, MA 01003, USA}

\author[0000-0001-8592-2706]{Alexandra Pope}
\affiliation{Department of Astronomy, University of Massachusetts, 710 North Pleasant Street, Amherst, MA 01003, USA}

\author[0000-0003-2919-7495]{Christina C. Williams}
\affil{Steward Observatory, 
University of Arizona, 
933 N. Cherry
Tucson, AZ 85721 USA}

\author{Peter R. M. Eisenhardt}
\affil{Jet Propulsion Laboratory, California Institute of Technology, 4800 Oak Grove Dr, Pasadena, CA 91109, USA}




\begin{abstract}

We present the average gas properties derived from ALMA Band~6 dust continuum imaging of 126 massive (log~$M_{\star}/\Msun\gtrsim10.5$), star-forming cluster galaxies across 11 galaxy clusters at $z=1-1.75$.  Using stacking analysis on the ALMA images, combined with UV-far-infrared data, we quantify the average infrared SEDs and gas properties (molecular~gas~masses, $M_{\rm mol}$; gas~depletion~timescales, $\tau_{\rm depl}$; and gas~fractions, f$_{\rm gas}$) as a function of cluster-centric radius and properties including stellar mass and distance from the Main Sequence.   We find a significant dearth in the ALMA fluxes relative to that expected in the field $-$ with correspondingly low $M_{\rm mol}$ and f$_{\rm gas}$ and short $\tau_{\rm depl}$ $-$ with weak or no dependence on cluster-centric radius out to twice the virial radius.  The Herschel+ALMA SEDs indicate warmer dust~temperatures ($\sim36-38$~K) than coeval field galaxies ($\sim30$~K).  We perform a thorough comparison of the cluster galaxy gas properties to field galaxies, finding deficits of 2-3x, 3-4x, and 2-4x in $M_{\rm mol}$, $\tau_{\rm depl}$, and f$_{\rm gas}$ compared to coeval field stacks and larger deficits compared to field scaling relations built primarily on detections.   The cluster gas properties derived here are comparable with stacking analyses in (proto-)clusters in the literature and at odds with findings of field-like $\tau_{\rm depl}$ and enhanced f$_{\rm gas}$ reported using CO and dust~continuum detections.  Our analysis suggests that environment has considerable impact on gas properties out to large radii, in good agreement with cosmological simulations which project gas depletion begins beyond the virial radius and largely completes by first passage of the cluster core.

\end{abstract}

\keywords{Galactic and extragalactic astronomy (563), Galaxy clusters (584), High-redshift galaxy clusters (2007), Molecular gas (1073), Galaxy quenching (2040)}


\section{Introduction} \label{sec:intro}

It has long been established that local environment and galaxy properties are linked, with overdense regions in the local Universe playing host to an overabundance of early type galaxies undergoing passive evolution \citep[e.g.][]{dre80}.  Substantial efforts have now been made to trace this relation back in time to the initial conditions of proto-clusters \citep[see][for a review]{ove16}, whose resident galaxies are predicted to contribute substantially to the cosmic star formation rate density and mass growth \citep{chi17}.  Bridging the gap between these early structures and present-day clusters is a transition epoch at $z\sim1-2$ wherein massive (log $M/\Msun\gtrsim13.8$) clusters are found to host populations of (dust obscured) star forming galaxies (SFGs) with field-like star formation activity \citep[i.e][hereafter A14, A16]{coo06, hil10, tra10, fas11, fas14, hay11, tad11, bro13, san14, san15, ma15, alb14, alb16} and a corresponding decrease in quenched populations and quenching efficiency \citep{nan17, cha19}.

Despite this progress, it is still unclear which of several mechanisms play a substantial role in quenching galaxies from the proto-cluster to cluster regimes.  Strangulation $-$ the prevention of fresh gas accretion due to the hot Intracluster Medium \citep[ICM; ][]{lar80} $-$ likely plays some role but its few Gyr timescales are not consistent with evidence supporting rapid quenching \citep[i.e.][A14, A16]{man10, vdb13, bro13, wet13}.  Ram pressure stripping \citep[RPS;][]{gun72} may be able to remove gas on shorter timescales \citep[$\sim100-200$ Myr;][]{aba99, mar03, roe06, roe07,kro08, ste16}.  RPS may also play a role out to large radii ($\gtrsim2-3R_{\rm vir}$) where infalling galaxies encounter virial accretion shocks associated with the ICM \citep[i.e.][]{sar98, bir03, dek06, zin18}. However, the effect of RPS on star formation (whether enhancement or quenching occurs) and gas reservoirs can depend on an individual galaxy's properties and orbit \citep{bek14, ton19}, making its overall effectiveness unclear.  Overdense environments may also host increased merger \citep{deg18, wat19} and AGN \citep[][A16]{mar13} activity, enabling scenarios in which quenching is propelled by increased starburst activity or feedback \citep{bro13}.

Key observables in distinguishing between these scenarios involve the cold molecular gas which fuels star formation: the (molecular) gas mass, gas depletion timescale, and gas fraction. In the local Universe, gas properties have been observed in the ever accumulating examples of RPS events, from truncated or disturbed gaseous disks \citep[i.e.][]{vol08, zab19} to spectacular one-sided tails \citep[i.e.][]{sun06, sun10, ebe14}, which can host significant star formation (SF) \citep{fum14} and molecular gas \citep{jac14, jac19}.  However it again remains unclear whether this cold gas is stripped or formed $in$ $situ$ in the tail and how this effects the star formation occurring in the host galaxy.  

At higher redshifts, where different mechanisms may operate, observations of gas content have been limited to small samples of cluster \citep[][Williams, ApJ, submitted]{nob17, nob19, rud17, sta17, hay17, hay18, coo18, spe21a, spe21b} and proto-cluster galaxies \citep{wan16, wan18, ume17, zav19, tad19, lon20, cha21, hil21} detected in CO or dust continuum emission.  These studies have largely found enhanced or field-like gas fractions and field-like gas depletion timescales, even when obtaining deep observations \citep[][Williams, et al., ApJ, submitted; but see \citet{coo18}]{nob19}, with only hints of gas loss at the highest stellar masses (log $M_{\star}/\Msun>11$).  On the other hand, the use of stacking to probe below detection limits suggests the existence of a population of cluster galaxies with low gas fractions and short depletion timescales \citep{bet19, zav19}.   Theoretically, evidence from cosmological simulations is mounting that gas loss should start at large radii, with significant depletion of the gas reservoir by or at first passage of the cluster center \citep{oma16, zin18, art19, mos21, oma21}.

In this work, we quantify the average molecular gas properties of cluster galaxies through stacking their dust continuum emission as observed by the Atacama Large Millimeter Array (ALMA).  Dust continuum emission, like CO line emission, is a robust proxy for molecular gas mass \citep[see][for a review]{tac20}. Our sample consists of 126 cluster galaxies selected, based on far-infrared (FIR) luminosity, from eleven stellar mass-selected galaxy clusters (log $M_{200}/M_{\odot}\sim14$) at $z=1-1.75$.  In general, the massive cluster galaxies in these clusters supply a total SFR budget comparable to massive field halos when controlled for halo mass, though there is significant variation in the total SF from cluster to cluster \citepalias{alb16}.  We measure the average gas properties of a luminous subset of these massive cluster galaxies, comparing their gas reservoirs to coeval field populations as well as to the (proto-)cluster samples in the literature at comparable redshifts. 

This paper is structured as follows: in Section~\ref{sec:data}, we describe our sample selection and observations.  In Section~\ref{sec:prop_stack}, we present a breakdown of the properties of our sample, our stacking techniques, and the methodology used to measure gas properties from our ALMA data.  Section~\ref{sec:analysis} contains our analysis and results: the average Herschel+ALMA infrared SEDs of our cluster galaxies, a cluster-centric radial analysis of gas properties, and a comparison to the gas properties of field galaxies and (proto-)cluster galaxies at comparable redshifts. In Section~\ref{sec:disc}, we discuss our results, including putting our findings in the context of recent cosmological simulations looking at gas loss in infalling cluster galaxies. Section~\ref{sec:conc} contains our conclusions. Throughout this work, we adopt concordance cosmology: ($\Omega_{\Lambda}$ , $\Omega_{\rm M}$ , $h$)=(0.7, 0.3, 0.7), a \citet{kro01} IMF, and the \citet{spe14} Main Sequence (MS).  ``log" refers to log$_{10}$. 

\section{Data}\label{sec:data}

\subsection{Sample}\label{sec:sample}

Our sample consists of 126 cluster galaxies selected from eleven massive (log $M_{200}/\Msun\sim14$) galaxy clusters at $z=1-1.75$ with uniquely deep {\it Herschel}/PACS imaging at 100 and 160 $\mu$m.  The clusters are drawn from the IRAC Shallow and IRAC Distant Cluster Surveys \cite[ISCS, IDCS;][]{eis08, sta12}, identified as near-infrared (stellar mass) overdensities in (RA, Dec, photometric redshift) space and confirmed via targeted spectroscopic follow-up \citep{sta05, sta12, els06, bro06, bro11, bro13, eis08, zei12, zei13}. We note that these eleven clusters were chosen for follow-up based on being significant overdensities and {\it not} on the basis of their star formation activity, which shows considerable variation from cluster-to-cluster. See \citetalias{alb16} and references therein for a detailed description of our cluster sample and {\it Herschel}/PACS imaging.

Cluster galaxy membership was established using spectroscopic redshifts (spec-$z$s) where available and photometric redshifts (photo-$z$s) otherwise, starting from a Spitzer/IRAC 4.5$\mu$m catalog and using full photo-$z$ probability distribution functions to identify a complete catalog of massive (log $M_{\star}/\Msun \geq 10.1$) cluster galaxies.  Herschel/PACS photometry was then extracted using the spectroscopic and IRAC catalogs as positional priors.  The infrared luminosity $L_{\rm IR}$[8-1000$\mu$m] for each member was calculated by scaling one of two templates from \citet{kir15} to the Herschel/PACS 100$\mu$m flux density; these templates are known to describe the average optical to FIR properties of massive cluster galaxies in the ISCS, IDCS \citepalias{alb16}.  For purely star forming galaxies, the template {\tt MIR0.0} was adopted.  For members with AGN activity as determined via SED fitting \citepalias[see \S\ref{sec:ancillary} and][]{alb16}, the template {\tt MIR0.5} and a correction factor were used to account for the contribution from AGN at shorter wavelengths.  The sample in this work was then selected as IR-bright ($L_{\rm IR}\geq5\e{11}\Lsun$) members within 2 Mpc of the cluster centers \citep[$\sim2$x the virial radius,][]{bro07}. The cluster centers are known to within $\sim15\arcsec$ ($\sim130\,$kpc), set by the pixel scale of the cluster detection maps \citep{gon19} and confirmed through comparisons with X-ray centroids (Garcia et al.~in preparation).

\subsection{ALMA Data}\label{sec:almadata}

Dust continuum emission was observed in Band 6 (centered at 1.3mm or 231 GHz) for our sample of 126 cluster galaxies over 125 pointings in ALMA Cycle 3 proposal 2015.1.00813.S (PI: Alberts).  The target positions were obtained from the IRAC counterparts, used as priors for Herschel source extraction. The sample was split into 6 Science Goals (for ease of scheduling) with four target rms sensitivities calculated based on two \citet{kir15} templates (discussed in the previous section), listed in Table~\ref{tbl:observations}.  The maximum 7.5 GHz bandwidth was requested over four 1875 MHz spectral windows.  Observations were taken in June 2016 with configurations C40-2 and C40-4.  Data reduction was performed with the Common Astronomy Software Application ver. 4.7.2 \citep[CASA;][]{mcm07} with preliminary reductions revealing that all sources are non-detections and that there are no serendipitous $>3\sigma$ detections in the maps.  As such, the final continuum maps were generated using the CASA task {\tt TCLEAN} with natural weighting, a $0.2\arcsec$ pixel size, and no deconvolution (cleaning).  {\it u,v} tapering was applied to produce maps with similar beamsizes (Table~\ref{tbl:observations}) as needed for stacking; a 260$k\lambda$ x 220$k\lambda$ taper was found to maintain sensitivity while coming close to $1\arcsec$ resolution, comparable to resolution of the IRAC priors.  The average rms sensitivities achieved are within $\sim10\%$ of requested; they and the beamsizes of the sample subsets are listed in Table~\ref{tbl:observations}.  We discuss why our targets are undetected in Section~\ref{sec:ir_sed}.

\begin{table*}
\centering
\caption{Summary of ALMA Observations}
\begin{tabular}{lcccc}
\hline
\hline
Sample & Number of & Requested rms & Average rms & Beamsize \\ 
& Pointings & [$\mu$Jy beam$^{-1}$] & [$\mu$Jy beam$^{-1}$] & [arcsec] \\
\hline
\hline
Science Goal 1a & 30 & 140 & 141.7 & 0.92 x 0.72 \\
Science Goal 1b & 29 & 140 & 136.0 & 0.96 x 0.75 \\
Science Goal 2a & 21 & 90 & 91.5 & 0.93 x 0.73 \\
Science Goal 2b & 22 & 90 & 79.8 & 1.0 x 0.67 \\
Science Goal 3 & 15 & 70 & 78.0 & 0.97 x 0.77 \\
Science Goal 4 & 8 & 50 & 56.5 & 0.92 x 0.75 \\
\hline
\label{tbl:observations}
\end{tabular}
\end{table*}

\subsection{Ancillary Data}\label{sec:ancillary}

For each ALMA target, UV through near-infrared photometry is available as described in \citet{chu14}, which was used to derive photo-$z$s and identify AGN as described in \citetalias{alb16}.
For ten of the clusters in this work, deep {\it Spitzer}/MIPS 24$\mu$m imaging was obtained to $3\sigma$ depths of $156\,\mu$Jy to $36\,\mu$Jy spanning $z=1$ to $z=1.75$, providing a uniform depth in $L_{\rm IR}$ of 3\e{11}\,\Lsun \citep{bro13}.  Deep  Herschel/PACS imaging at 100 and 160$\mu$m (average rms sensitivity of $1.2\,\mu$Jy at 100$\mu$m) was obtained for all 11 clusters, as described in \citetalias{alb16}. Herschel/SPIRE imaging at 250, 350, and 500$\,\mu$m is available from the Herschel Multi-tiered Extragalactic Survey \citep[HerMES;][]{oli12} for 10/11 clusters; this work uses the SPIRE maps as reduced and described in \citetalias{alb14}.

\section{ALMA Sample Properties and Stacking}\label{sec:prop_stack}

\begin{figure*}
    \centering
    \includegraphics[width=\textwidth]{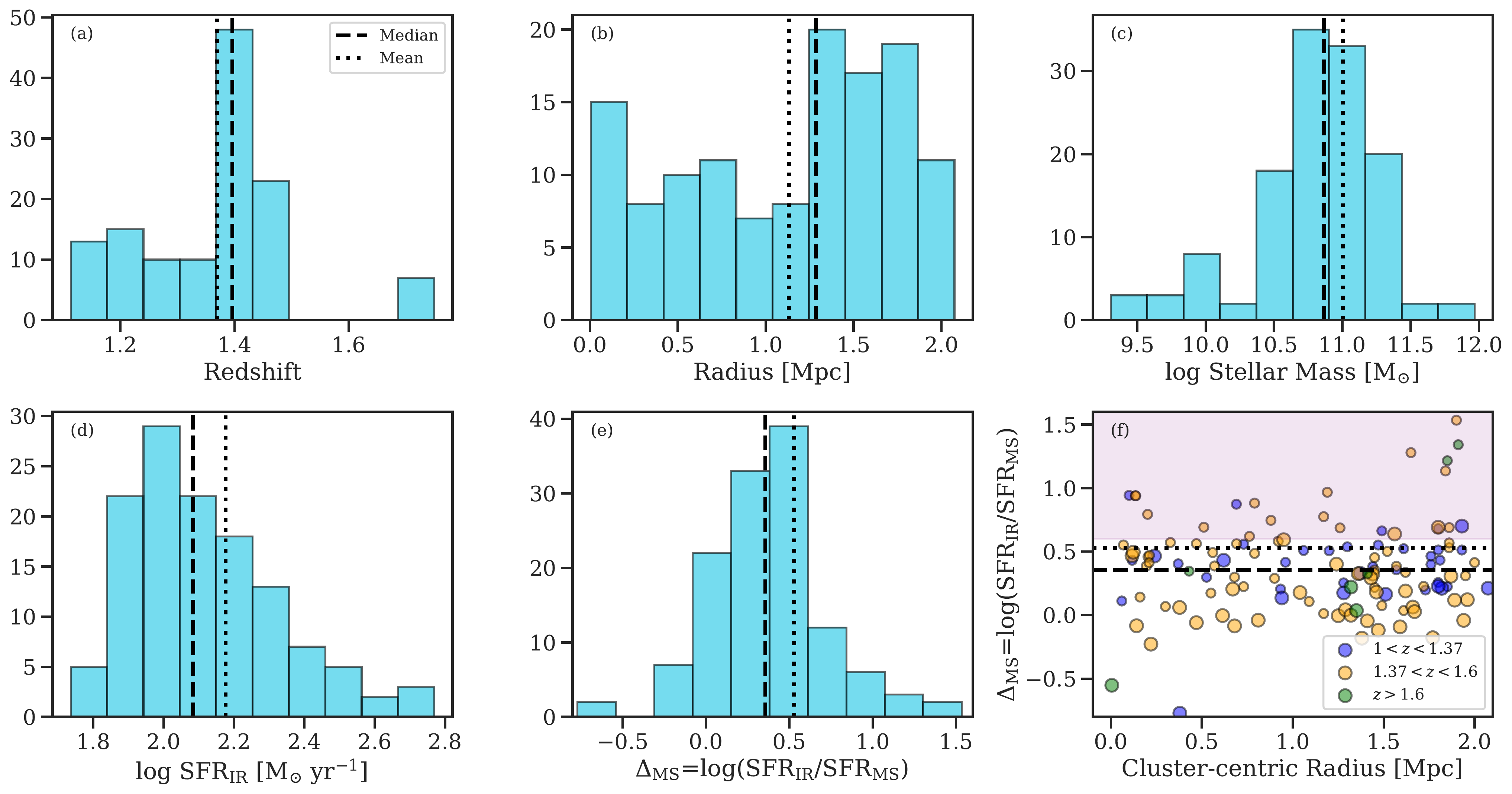}
    \caption{Properties of the ALMA cluster member sample: (a) redshift, (b) cluster-centric radius, (c) stellar mass, (d) obscured SFR, (e) $\Delta$MS. (f) shows the distance from the main sequence as a function of cluster-centric radius in three redshift bins, with the lower and upper half of our stellar mass distribution indicated by smaller and larger symbols. Median and mean values are indicated as dashed and dotted lines. The purple shaded region denotes starbursting activity.}
    \label{fig:properties}
\end{figure*}

Histograms of the following properties of our ALMA sample are shown in Figure~\ref{fig:properties}: redshift (spec-$z$ or photo-$z$), cluster-centric (projected) radius, stellar mass, obscured star formation rate SFR$_{\rm IR}$, and distance from the Main Sequence.  The photo-$z$ uncertainties are $\sigma/(1+z) \approx 0.05$ for sources dominated by (host) galaxy emission and $\sigma/(1+z) \approx 0.2$ when dominated by AGN emission \citepalias[see][]{alb16}, based on comparisons with spectroscopic redshifts and pair statistics \citep{qua10, hua13}.  Stellar masses were derived using optical - MIR Bayesian SED fitting \citep{mou13} as described in \citet{bro13}, with uncertainties of 0.3 dex including systematic error. SFR$_{\rm IR}$ was obtained from the total $L_{\rm IR}$ (as described in \citetalias{alb16} and in Section~\ref{sec:data}) determined using Herschel/PACS photometry and assuming the $L_{\rm IR}$-SFR conversion from \citet{mur11b}, with corrections for AGN emission where appropriate \citep{kir15}.  The typical (median) measurement uncertainty on SFR$_{\rm IR}$ is $\sim0.15$ dex \citepalias[][]{alb16}.  We note that for massive galaxies at these redshifts, the obscured component contributes the bulk of the star formation rate \citep{whi17}.  The Main Sequence star formation rate (SFR$_{\rm MS}$) for each of our targets was determined from their redshifts and stellar masses using the publicly available Python package {\tt a3cosmos} \citep{liu19a} and the \citet{spe14} MS relation.  \citet{spe14} assumes a \citet{cha03} IMF, which has a similar normalization to the \citet{kro01} IMF assumed in our $L_{\rm IR}$ to SFR conversion \citep{mur11b}.  From this we derived the distance of our sources from the MS as $\Delta \mathrm{MS}=\mathrm{log}(\mathrm{SFR}_{\rm IR}/\mathrm{SFR}_{\rm MS})$.  

To summarize (Figure~\ref{fig:properties}, (a)-(e)), our sample is largely massive (log $M_{\star}/\Msun\sim11$) cluster galaxies at $z\sim1.4$ with SFRs on the high end of the MS at this redshift with some starbursting activity.  
We cover up to 2x the virial radius \citep[$R_{\rm vir}\sim1\,$Mpc,][]{bro07} and sit above the MS with a mean (median) $\Delta$MS of 0.52 (0.36), with 23 members falling into the category of starbursts \citep[$\Delta$MS$\,\geq0.6$;][]{rod11, rod14}.  Figure~\ref{fig:properties} (f) shows that the starbursts are not preferentially located at any particular redshift or cluster-centric radius, though they tend to be in the lower half of our stellar mass distribution. This is due to our flux-limited selection, which results in a bias with stellar mass; at lower masses (log $M_{\star}/\Msun\lesssim11$), we only probe sources on the upper end and above the MS with log $M_{\star}/\Msun\lesssim10.3$ sources making up the majority of our starbursts.

\subsection{Stacking and Stacked Photometry}

\subsubsection{ALMA}\label{sec:almastacking}

Stacking is performed by creating 31x31 pixel cutouts of the primary-beam corrected maps centered on the target coordinates.  Except for two cluster galaxies, all of our targets are at the center of the ALMA pointings.  Cutouts are then binned into subsets by different properties and the pixel-wise variance-weighted mean is calculated.  The variance-weighted mean provides the most robust average flux density in the case of non-uniform noise properties when combining cutouts that reach different depths (Table~\ref{tbl:observations}).

\begin{figure*}
    \centering
    \includegraphics[width=\linewidth, trim=10 270 0 200, clip]{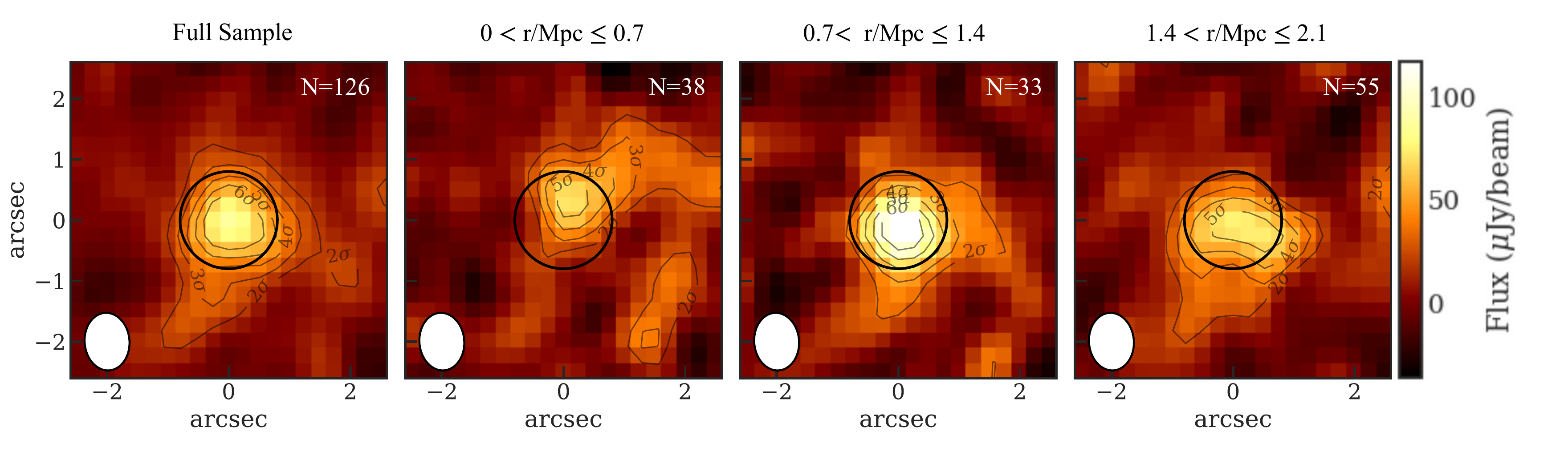}
    \caption{Weighted-mean stacks at (observed) 1.3mm for our full cluster sample (far left) and split into three cluster-centric radial bins.  The black circle shows the aperture used to measure the stacked photometry (see Section~\ref{sec:almastacking}).  The average beamsize is shown in the lower left corner.}
    \label{fig:stacks}
\end{figure*}

Figure~\ref{fig:stacks} shows the stack of the full sample, which is marginally resolved. This is consistent with the median size of the (global) star forming regions observed in MS galaxies at $z\sim2$ with deep ALMA and VLA radio data \citep[$r_{\rm SF}\sim2.1\pm0.9$ kpc;][]{ruj16}.  Accordingly, we use integrated aperture photometry to determine the stacked flux following the procedure used in \citet{bet19}: starting at the average (circularized) beamsize, the S/N is measured in increasingly large apertures in order to identify the aperture which maximizes the S/N by enclosing the most signal while minimizing the contribution from noise (determined off source as the standard deviation of 100 apertures).  From this procedure, we find that the optimal aperture is $0.8\arcsec$ in radius\footnote{We confirm that the aperture photometry returns a larger flux density than the peak pixel, which provides a measure of the source flux in images calibrated in Jy beam$^{-1}$ in the case of a point source.}.  We use this aperture across all of our stacks.

Photometric uncertainties are derived using the bootstrap technique \citep[i.e.][]{bet12} as described in \citetalias{alb14}, which encompasses all sources of noise in the images as well as scatter in the ALMA properties of the stacked population.  Stacking and aperture photometry are performed on randomly selected cutouts with replacement for $N=500$ realizations; the stacked uncertainty is then measured from the width of the resulting histogram.

\subsubsection{SPIRE}\label{sec:spirestacking}

Stacking in the SPIRE 250, 350, and 500$\,\mu$m bands was performed as described above, on cutouts centered on the target galaxies to determine the variance-weighted mean.  Given the large SPIRE beamsizes (18, 25, and 36\arcsec at 250, 350, and 500$\,\mu$m) and the high source density of the clusters, these stacks will suffer from flux boosting \citep{vie13}.  While it is possible to correct for this flux boosting statistically in large samples \citep[A14, ][]{alb21}, we utilize simulations \citep[see Appendix~C in ][]{alb21} to determine that the correction has a large uncertainty in small samples like those being stacked in this work, particularly at the longer wavelengths.  As such, we adopt the uncorrected SPIRE stacks, which are formally upper limits.  We note that in a previous study flux boosting at 250$\,\mu$m was found to sharply decrease with cluster-centric radius and is minimal at $r>0.5\,$Mpc for this cluster sample \citepalias{alb14}.   Stack uncertainties are determined via bootstrapping.

\subsection{Molecular Gas Measurements}\label{sec:gasmeasurements}

Over the redshift range of our clusters ($z\sim1-1.75$), our observed ALMA 1.3mm photometry probes rest-frame 650-470$\mu$m on the optically thin Rayleigh Jeans tail of the dust emission.  This provides a robust measure of the total dust mass; from this, given assumptions on the dust opacity and dust-to-gas abundance ratio, the total molecular gas mass can be derived \citep{eal12, sco14, pri18}.  This method agrees well with gas measurements from CO emission lines, see \citet{tac20} for a review.

All dust mass and gas mass measurements used in this work are calculated following \citet{sco16}, hereafter S16.  We briefly summarize the method here: in the optically thin regime, the observed flux density, $S_{\nu}$, is a function of the dust opacity per unit mass, $\kappa_{\nu}^{dust}$, the mass-weighted dust temperature T$^{\rm bulk}_{\rm{dust}}$, and the dust mass M$_{\rm{dust}}$ though 

\begin{equation}
\label{eqn1}
S_{\nu} = \kappa_{\nu}^{\mathrm{dust}} \mathrm{T^{bulk}_{dust}} (1+z)  \nu^2 \frac{\mathrm{M_{dust}}}{4\pi d_L^2}
\end{equation}
where $d_L$ is the luminosity distance. This can be related to the total molecular gas mass, $M_{\rm mol}$, by defining the dust opacity per unit ISM mass via $\kappa^{\rm{mol}}_{\nu} = \kappa^{\rm{dust}}_{\nu} \times \rm{M_{dust}/M_{mol}}$.   The gas mass can then be quantified given photometry on the RJ tail of the dust emission, the dust temperature, and $\kappa^{\rm{mol}}_{\nu}$.

$\kappa^{\rm{mol}}_{\nu}$ was empirically calibrated at 850$\mu$m based on local SFGs and Ultra-Luminous Infrared Galaxies (ULIRGs) as well as high-$z$ Sub-millimeter Galaxies \citep[SMGs;][]{sco14, sco16} and $z\sim2$ SFGs \citep{kaa19}, from which was found a single representative constant:

\begin{equation}
\label{eqn2}
\alpha_{850} \propto \kappa^{\mathrm{mol}}_{\nu_{850}} \mathrm{T^{bulk}_{dust}} = \frac{L_{\nu_{850}}}{\mathrm{M_{mol}}} = 6.7\e{19} \mathrm{ergs\,} \mathrm{s}^{-1} \mathrm{Hz}^{-1} \Msun^{-1}. 
\end{equation}

assuming a bulk dust temperature of $T^{\rm bulk}_{\rm dust}=25$ K \citep{kir15,sco16}.  We apply this calibration to derive $M_{\rm mol}$, accounting for redshift, our observed frequency, and departures from Rayleigh Jeans as in Eqn. 16 of \citetalias{sco16} (please note the typo as corrected in their erratum).

\section{Analysis and Results}\label{sec:analysis}

\subsection{Infrared SEDs}\label{sec:ir_sed}

In \citetalias{alb16}, we showed that up to the dust peak ($\sim$80-100$\,\mu$m), the infrared SED of massive, star forming cluster galaxies at $z\sim1-2$ is well represented by SFG templates derived from field galaxies in the same epoch.  Here we extend that analysis to the submm by adding to the SED our stacked ALMA fluxes in three bins of cluster-centric radius: $0<r/\mathrm{Mpc}<0.7$, $0.7<r/\mathrm{Mpc}<1.4$ and $1.4<r/\mathrm{Mpc}<2.1$.  
These radial bins were chosen in part to ensure good detections in the stacks. We can interpret them physically as roughly representing different environments within the cluster ecosystem via the following: simulated mass profiles have found a sharp drop in cluster dark matter profiles, termed the splashback radius \citep[$R_{\rm sp}$;][]{die14}, which denotes the boundary between the virialized and infalling regions.  $N$-body simulations and observations (via weak lensing) both agree that, for rapidly accreting halos such as we expect at high redshifts, the splashback radius roughly coincides with the virial radius \citep[i.e.][and references therein]{mor15, die17b, zur19, shi19, shi21}.  Though we do not have the data to measure the splashback radii individually in our clusters, this is qualitatively confirmed by stacked stellar mass profiles of the full ISCS cluster sample, which showed a sharp drop off at the virial radius \citep{alb21}.  As such, our data out to $2R_{\rm vir}$ is likely allowing us to separately probe the virialized and infalling populations, as well as the transition region (at $\sim R_{\rm vir}$) between the two.
The three radial stacks can be seen in Figure~\ref{fig:stacks}; all radial stacks are detected at S/N$\,\sim4$ (Table~\ref{tbl:properties}).

\begin{figure*}
    \centering
    \includegraphics[width=\linewidth]{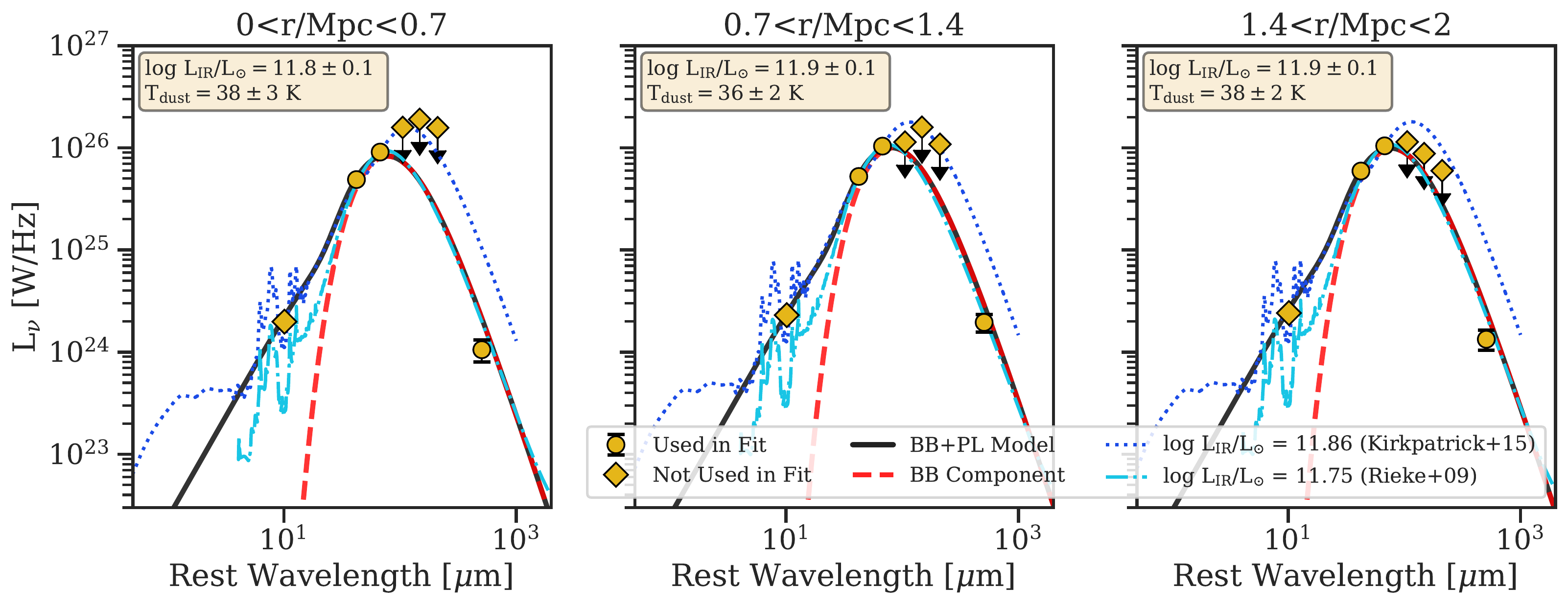}
    \caption{The IR SEDs of our targets divided into three cluster-centric radial bins: $0<r/\mathrm{Mpc}<0.7$, $0.7<r/\mathrm{Mpc}<1.4$, $1.4<r/\mathrm{Mpc}<2.1$, at the mean redshifts of these bins (Table~\ref{tbl:properties}).  The black line shows a fit to the average flux densities at 100, 160, and 1300$\mu$m (yellow circles), parameterized as a MIR power-law plus a modified blackbody.  The median MIPS 24$\mu$m and stacked SPIRE 250, 350, and 500$\,\mu$m (uncorrected for flux boosting, see Section~\ref{sec:spirestacking}) flux densities are shown (yellow diamonds), but not included in the fit.  The dashed red line shows the modified blackbody component only.  We compare to two field-based SFG templates: the log $L_{\rm IR}/\Lsun\sim11.75$ local template from \citet{rie09} and the $z\sim2$ template from \citet{kir15}, normalized to the 160$\mu$m datapoint.  We find that the local template, which has a warm dust peak and relatively weak submm emission, better represents our cluster galaxies on average, with no apparent radial dependence.}
    \label{fig:seds}
\end{figure*}

To quantify the IR SEDs (Figure~\ref{fig:seds}) in these radial bins, we combine the average flux densities from the detections of our sources at (observed) 100 and $160\mu$m with the stacked average flux density at 1.3mm.  Shorter wavelength MIPS 24$\mu$m detections are available for a subset of our sample; however, these probe the complex PAH and absorption features in the MIR, which will be diluted due to the redshift range probed.  Accordingly, we display the median value of the 24$\,\mu$m flux densities available but do not include these in our SED fitting.  We also display the SPIRE stacks at 250, 350, and 500$\,\mu$m; however, as discussed in \S~\ref{sec:ancillary} and \S~\ref{sec:spirestacking}, SPIRE coverage is only available for 10/11 clusters and the small number of sources stacked prevents us from applying a robust correction for flux boosting.  Accordingly, the SPIRE stacks are shown as upper limits and not included in the fitting.  The flux boosting is maximal for the largest beamsize (36\arcsec at 500$\,\mu$m) and at small cluster-centric radius.  It is expected to be minimal at $r>0.5\,$Mpc at 250$\,\mu$m \citepalias{alb14}.

Following \citet{cas12}, we parameterize the dust emission as a mid-IR power law arising from warm/hot dust emission from compact star forming regions and/or AGN plus a single-temperature modified blackbody, representing the cold dust emission from reprocessed light from young stars.  The dust emissitivy index ($\beta$=1.5) and MIR power law index ($\alpha$=2) are fixed, and general opacity is assumed, appropriate for conditions near the peak of emission.  Total $L_{\rm IR}$ and $T_{\rm dust}$ (i.e. the dust temperature representing the luminosity-weighted dust emission) are allowed to vary.  

The addition of the submm point to the data at the FIR peak indicates that that the full FIR SED is warmer ($\sim$36-38 K) than typical of massive field galaxies at this redshift \citep[$\sim30$ K;][]{sch18}, with weaker submm emission relative to the dust peak than found for the $z\sim1$ SFG template that well represents luminous field ALMA sources in blank field surveys \citep{kir15, dun17}. Given the degeneracy between dust temperature and emissivity index, we repeat this analysis assuming $\beta=2$ \citep{wei01}, finding lower dust temperatures by $\sim2$ K.  As \citet{sch18} assumes $\beta=1.5$, we cautiously present our $\beta=1.5$ dust temperatures as the direct comparison to the field value.  It is unknown at this time whether $\beta$ varies as a function of environment. The average total infrared luminosity in all three radial bins is log $L_{\rm IR}/\Lsun\approx11.9$.  We note that the MIPS 24$\,\mu$m is in good agreement with the fit and that the SPIRE fluxes are consistent with warmer dust temperatures where flux boosting is expected to be minimal, in the outermost radial bin. We discuss these warm dust temperatures in the context of the literature on galaxy clusters in \S~\ref{sec:warm_dust}. 

A recent compilation of targeted ALMA observations of galaxies at $z\sim2-4$ with robust photometry spanning the dust peak found that local templates from \citet{rie09} can reproduce the full FIR SED at these redshifts \citep{der18}.  In Figure~\ref{fig:seds}, we compare to the \citet{kir15} and \citet{rie09} templates which most closely match our measured $L_{\rm IR}$, normalized (not fit) at 160$\mu$m; we find that the local \citet{rie09} template better reproduces our cluster SEDs\footnote{We note that we do not update our total $L_{\rm IR}$ and SFR$_{\rm IR}$ measurements as described in \S~\ref{sec:sample} to use the \citet{rie09} template as these quantities are heavily luminosity-weighted. At these redshifts and based on Herschel/PACS measurements near the dust peak, the expected difference in $L_{\rm IR}$ calculated using \citet{kir15} versus \citet{rie09} templates is on order $\lesssim30\%$.}. We confirm via least squares minimization that the \citet{rie09} template has reduced $\chi^2$ values of a few, an order of magnitude lower than the reduced $\chi^2$ of the \cite{kir15} template. Predicting the submm from the FIR peak and field templates such as those found appropriate in \citet{kir15} and \citet{dun17} overestimates the average ALMA flux of our targets by a factor $\sim6.5$, resulting in non-detections for observations aimed at moderate S/N.

\begin{table*}
\renewcommand{\arraystretch}{1.15}
\centering
\caption{Stacked ALMA emission and average derived properties of subsets of cluster galaxies. }
\begin{tabular}{lcccccccccc}
\hline
\hline
Subset & Number in & Stacked       & S/N & $z$ & log $M_{\star}$ & SFR$_{\rm IR}$ & $\Delta$MS & log $M_{\rm mol}$ & $\tau_{\rm depl}$ & $f_{\rm gas}$ \\ 
       & Stack     & $S_{1.3}$ [$\mu$Jy] &     &  &  [\Msun]    &  [\Msun yr$^{-1}$] &    &  [$\Msun$]     &   [Gyr]                & \\
\hline
\hline
All & 126 & 123.1$\pm$17 & 7.4 & $-$ & $-$ & $-$ & $-$ & $-$ & $-$ & $-$ \\
\hline
\multicolumn{10}{c}{Split by Radius}\\
\hline
$0<r/\mathrm{Mpc}<0.7$ & 38 & 90.1$\pm$21 & 4.3 & 1.38 & 11.1$\pm$0.1 & 129$\pm$6 & 0.17 & 10.2$\pm$0.1 & 0.11$\pm$0.03 & 0.11$\pm$0.3 \\
$0.7<r/\mathrm{Mpc}<1.4$ & 33 & 158.6$\pm$36 & 4.4 & 1.37 & 11.0$\pm$0.1 & 149$\pm$7 & 0.29 &  10.4$\pm$0.1 & 0.17$\pm$0.04 & 0.20$\pm$0.05 \\
$1.4<r/\mathrm{Mpc}<2.1$ & 55 & 114.2$\pm$ 30 & 3.8 & 1.37 &  11.0$\pm$0.1 & 165$\pm$6 & 0.33& 10.3$\pm$0.1 & 0.11 $\pm$0.03 & 0.16$\pm$0.05\\
\hline
\multicolumn{10}{c}{Split by Median Property Value}\\
\hline
$z<1.396$ & 59 & 106.5$\pm$23 & 4.6 & 1.26 & 11.0$\pm$0.06 & 138$\pm$5 & 0.30 & 10.2$\pm$0.09 & 0.12$\pm$0.03 & 0.14$\pm$0.04\\
$z>1.396$ & 67 & 130.6$\pm$25 & 5.2 & 1.47 & 11.0$\pm$0.05 & 161$\pm$5 & 0.26 & 10.3$\pm$0.08 & 0.13$\pm$0.02 & 0.17$\pm$0.04 \\
\hline
log $M_{\star}/\Msun<10.87$ & 68 & 110.0$\pm$23 & 4.8 & 1.34 & 10.7$\pm$0.04 & 128$\pm$5 & 0.42 & 10.2$\pm$0.09 & 0.14$\pm$0.03 & 0.24$\pm$0.05 \\
log $M_{\star}/\Msun>10.87$ & 48 & 166.5$\pm$27 & 6.3 & 1.39 & 11.3$\pm$0.05 & 181$\pm$6 & 0.16 & 10.4$\pm$0.07 & 0.15$\pm$0.02 & 0.12$\pm$0.02 \\
\hline
SFR$_{\rm IR}<121\,\Msun$ yr$^{-1}$ & 62 & 92.5$\pm$20 & 4.6 & 1.35 & 11.0$\pm$0.07 & 90$\pm$5 & 0.09 &  10.2$\pm$0.09 & 0.16$\pm$0.04 & 0.13$\pm$0.03\\
SFR$_{\rm IR}>121\,\Msun$ yr$^{-1}$ & 64 & 193.6$\pm$29 & 6.7 & 1.39 & 11.0$\pm$0.05 & 208$\pm$5 & 0.39 & 10.5$\pm$0.06 & 0.15$\pm$0.02 & 0.22$\pm$0.04 \\
\hline
$\Delta$MS$\,<0.35$ & 62 & 133.4$\pm$24 & 5.5 & 1.40 & 11.2$\pm$0.05 & 125$\pm$5 & 0.07 & 10.3$\pm$0.08 & 0.17$\pm$0.03 & 0.12$\pm$0.03\\
$\Delta$MS$\,>0.35$ & 64 & 97.1$\pm$26 & 3.7 & 1.34 & 10.7$\pm$0.05 & 174$\pm$5 & 0.55 & 10.2$\pm$0.12 & 0.09$\pm$0.02 & 0.22$\pm$0.06\\
\hline
\hline
\label{tbl:properties}
\end{tabular}
\end{table*}

\subsection{Molecular gas masses, gas depletion time scales, and gas fractions}

Molecular gas masses are measured following \citetalias{sco16} as described in Section~\ref{sec:gasmeasurements}.  From these, we can quantify two key relations between gas content and other galaxy properties.  First, the gas depletion time ($\tau_{\rm depl}\equiv M_{\rm mol}/\mathrm{SFR}$), or its inverse the star formation efficiency (SFE), gives the timescale for the depletion of the molecular gas in the system assuming the current SFR and no new gas accretion or gas recycling.  Second, the gas mass fraction ($f_{\rm gas}\equiv M_{\rm mol}/(M_{\star} + M_{\rm mol})$) relates the current gas content to the stellar mass.

In \citetalias{alb16}, we analyzed the star forming properties of the IR luminous ISCS/IDCS cluster galaxies from which the sample in this work is drawn \citep[see also][A14]{bro13}.  We found evidence for a rapid transition epoch in our mass-limited SFG population at $z\sim1.4$; on average, cluster galaxies have SFRs comparable to field galaxies at this epoch, with evidence for rapid quenching seen in the sharp decrease in the star forming fraction and average specific-star formation rate (SSFR) across a relatively short timescale of our redshift range ($\lesssim1$ Gyr). Evidence for rapid quenching at $z\sim1.5$ has similarly been found by looking at the environmental quenching efficiency via quiescent cluster populations \citep{nan17}.   

\begin{figure}
    \centering
    \includegraphics[scale=0.55, trim=5 0 5 0, clip]{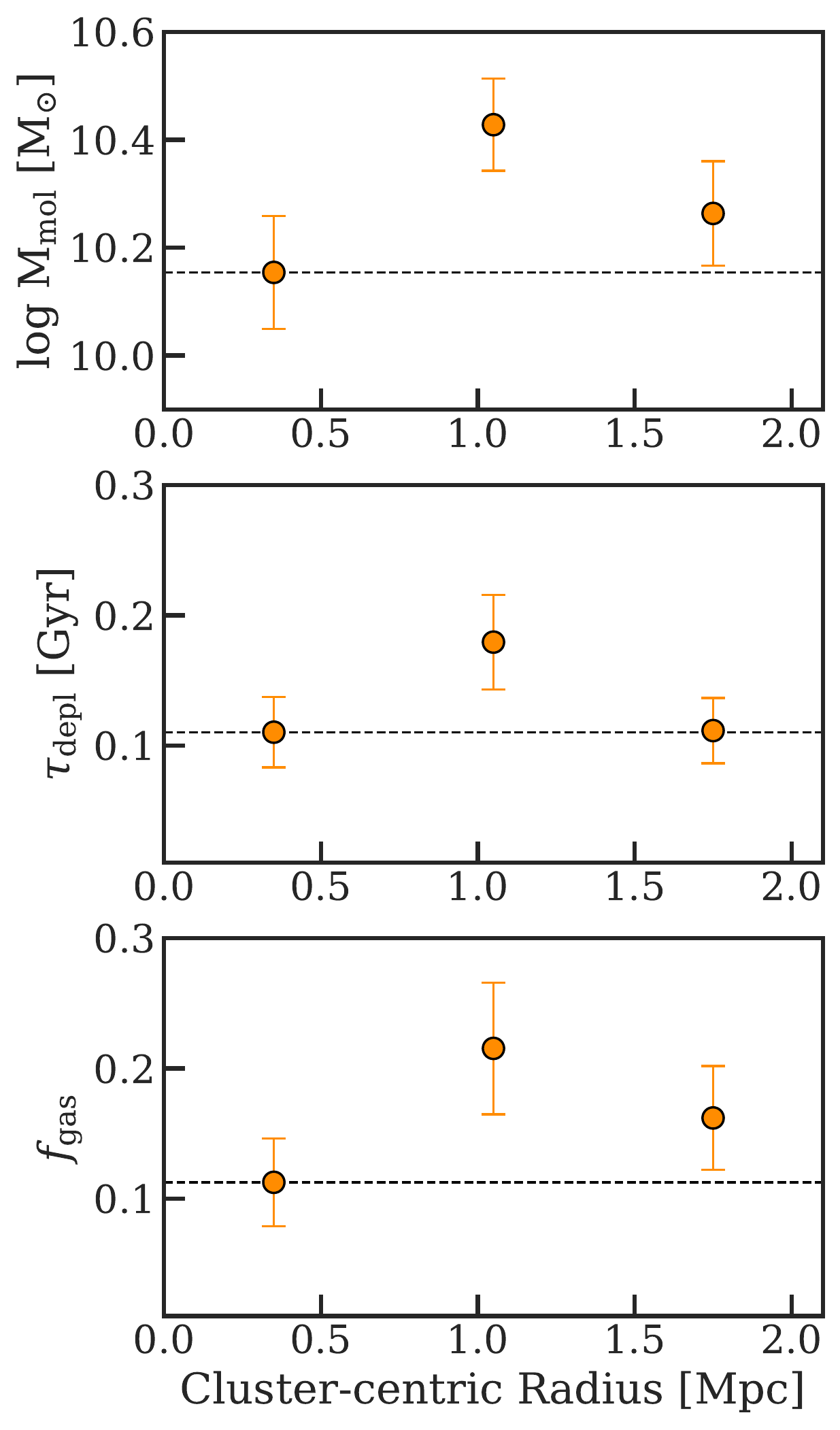}
    \caption{The molecular gas mass (top), gas depletion timescale (middle) and gas fraction (bottom) of cluster galaxies in three projected radial bins: $0<r/\mathrm{Mpc}<0.7$, $0.7<r/\mathrm{Mpc}<1.4$ and $1.4<r/\mathrm{Mpc}<2.1$.  Stacked uncertainties are from bootstrapping and so incorporate the spread in the stacked populations.  The dashed line shows the level of the cluster core, to guide the eye.  There is weak to no dependence of these properties on cluster-centric radius out to twice the virial radius ($R_{\rm vir}\sim1\,$ Mpc).}
    \label{fig:radial}
\end{figure}

Here we examine the gas content of our ALMA targets, cluster SFGs on the high end of the MS.  In Figure~\ref{fig:radial}, we show the average $M_{\rm mol}$, $\tau_{\rm depl}$, and $f_{\rm gas}$ for the three radial bins analyzed in the previous section: $0<r/\mathrm{Mpc}<0.7$, $0.7<r/\mathrm{Mpc}<1.4$ and $1.4<r/\mathrm{Mpc}<2.1$.  Again, assuming $R_{\rm sp}\sim R_{\rm vir}$, these bins are probing the virialized, transition, and infalling regions of our clusters.  We find that on average, our cluster galaxies have low molecular gas masses (log $M_{\rm mol}/\Msun\sim10.3$) and correspondingly low gas depletion times ($\sim100-200\,$Myr) and gas fractions (0.1-0.2).  We compare these values to their field counterparts in the next section.  We find only a weak radial dependence; cluster galaxies near the virial radius show slightly elevated gas content and longer gas depletion timescales than those in the cluster cores or well outside the virial (and potentially splashback) radius.  This transition region is likely dominated by galaxies entering the virialized region for the first time, as the population of backsplash galaxies (cluster galaxies that have already fallen into and past the cluster core, whose orbits are again on outbound trajectories) is  expected to be small at this redshift \citep{mos21}.  However, this difference is on the $1\sigma$ level when accounting for the bootstrapped errors which incorporate the spread in the population being stacked.  Overall, the gas content is low across all radii, agreeing with previous observational works that show the cluster influence extending to at least two times the virial radius \citep[i.e][]{bal99, vdl10, chu11, ras12, shi21}.

We note two sources of potential contamination in this analysis.  The first is that we have identified the majority of our cluster galaxies using photometric redshifts, which will mis-classify some field galaxies as cluster galaxies.  This will serve to dilute signals of environmental effects.

The second is that we have used projected cluster-centric radii to create our bins, which will dilute any signal as a function of radius.  The use of phase space diagrams \citep[i.e.][]{rhe17} can place cluster galaxies in the context of their accretion history, and roughly separate virialized, infalling, and backsplash populations.  However, the spectroscopy needed has thus far been largely limited to 1) the optical, which can miss the heavily obscured cluster galaxies dominating the SF budget \citepalias{alb16}, or 2) cluster galaxies with high gas content through CO emission.  We argue in the next sections that the latter are not representative.  Sensitive near-infrared spectroscopy, such as provided by the {\it James} {\it Webb} {\it Space} {\it Telescope}, is needed for complete, mass-limited spectroscopic cluster catalogs.

\subsection{Comparison with the field: significant gas deficits in cluster SFGs}\label{sec:field_compare}

In the last section, we determined that our cluster galaxies generally have low gas masses and gas fractions, with short depletion timescales with weak or no dependence on cluster-centric radius.  In this section, we redo our stacking analysis by splitting our sample into subsets by the median values (Figure~\ref{fig:properties}, Table~\ref{tbl:properties}) of the following properties: stellar mass, redshift, SFR$_{\rm IR}$, and $\Delta$MS. We then compare to field galaxies in order to quantify environmentally-driven differences in gas properties.

\begin{figure*}
    \centering
    \includegraphics[width=\linewidth]{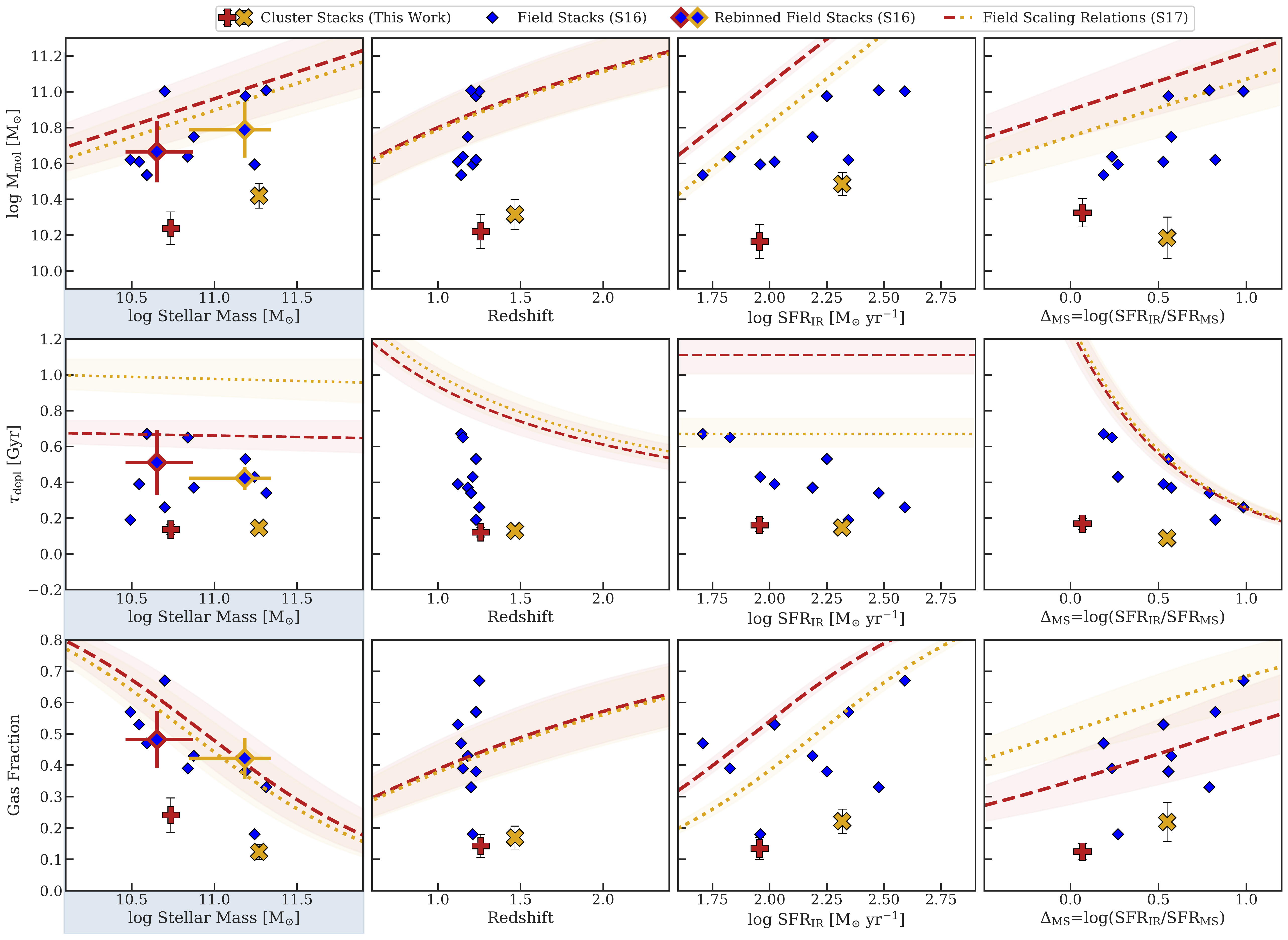}
    \caption{The average molecular gas masses (first row), gas depletion timescales (second row) and gas fractions (third row) as a function of galaxy properties: stellar mass (first column), redshift (second column), IR SFR (third column) and distance from the MS (fourth column).   In each panel, we measure the gas content from our stacks by splitting our ALMA sample into galaxies below (red cross) and above (yellow x) the median value of each property as listed in Table~\ref{tbl:properties}. The gas properties of field galaxy stacks \citetalias{sco16} are shown as blue diamonds.
    In the first column, the \citetalias{sco16} stacks are additionally rebinned into a weighted average in two mass bins for a more direct comparison. The red dashed and yellow dotted lines are the predicted values in each bin from field scaling relations \citepalias{sco17}, calculated using the average properties (redshift, stellar mass, $\Delta$MS) of the cluster stacks, as described in Section~\ref{sec:field_compare}.  Table~\ref{tbl:deficits} gives the cluster deficits as the ratio between the field and cluster values, using the rebinned \citetalias{sco16} stacks and \citetalias{sco17} scaling relations.  }
    \label{fig:field_compare}
\end{figure*}

\begin{table*}
\centering
\caption{The average deficits measured in gas mass, depletion timescale, and gas fraction obtained by taking the ratio of the predicted field value to the stacked cluster value (Figure~\ref{fig:field_compare}).  Deficits are calculated in comparison to the rebinned field stacks presented in S16 and the field scaling relations presented in S17 (see Section~\ref{sec:field_compare}).}
\begin{tabular}{lcccccc}
\hline
\hline
Subset & $M_{\rm mol}$  & $\tau_{\rm depl}$ & $f_{\rm gas}$ & $M_{\rm mol}$  & $\tau_{\rm depl}$ & $f_{\rm gas}$\\ 
       & S16$^\dagger$ & S16$^\dagger$ & S16$^\dagger$ & S17 & S17 & S17 \\
\hline
\hline
log $M_{\star}/\Msun<10.87$ & $2.7\pm1.0$ & $3.8\pm1.6$ & $2.0\pm0.6$ & $4.4^{+2.1}_{-1.7}$ & $4.9^{+1.2}_{-1.2}$ & $2.4^{+0.7}_{-0.7}$ \\
log $M_{\star}/\Msun>10.87$ & $2.3\pm0.9$ & $2.9\pm0.7$ & $3.4\pm0.8$ & $3.6^{+1.8}_{-1.3}$ & $6.7^{+1.3}_{-1.3}$ & $2.7^{+0.8}_{-0.9}$  \\
\hline
$z<1.396$ & $-$ & $-$ & $-$ & $4.7^{+2.3}_{-1.8}$ & $6.8^{+1.7}_{-1.7}$ & $3.1^{+1.0}_{-1.0}$   \\
$z>1.396$ & $-$ & $-$ & $-$ & $4.4^{+2.1}_{-1.6}$ & $6.2^{+1.4}_{-1.4}$ & $2.8^{+0.8}_{-0.8}$   \\
\hline
SFR$_{\rm IR}<121\,\Msun$ yr$^{-1}$ & $-$ & $-$ & $-$ & $4.9^{+2.2}_{-1.8}$ & $6.9^{+1.7}_{-1.7}$ & $3.2^{+1.0}_{-1.1}$   \\
SFR$_{\rm IR}>121\,\Msun$ yr$^{-1}$ & $-$ & $-$ & $-$ & $3.1^{+1.6}_{-1.1}$ & $4.6^{+0.9}_{-0.9}$ & $2.1^{+0.6}_{-0.6}$  \\
\hline
$\Delta$MS$\,<0.35$ & $-$ & $-$ & $-$ & $4.0^{+1.8}_{-1.4}$ & $6.7^{+1.4}_{-1.4}$ & $2.9^{+0.9}_{-0.9}$  \\
$\Delta$MS$\,>0.35$ & $-$ & $-$ & $-$ & $5.5^{+2.9}_{-2.3}$ & $6.1^{+1.8}_{-1.8}$ & $2.8^{+0.9}_{-0.9}$  \\
\hline
\hline
\label{tbl:deficits}
\end{tabular} \\
$^\dagger$Deficits from the S16 field are only calculated for the cluster stack subsets split by stellar mass.  See Section~\ref{sec:field_compare}.
\end{table*}

Ideally, the comparison of the gas content in cluster and field galaxies would be between samples well matched in terms of target selection criteria, stellar mass range, and the detection limits and methodology used to probe the gas.  Stacking analyses, as used in this work, have the advantage of not relying on individual detections, which can bias a study toward the gas-rich end.  Given these considerations, we begin our comparison in Figure~\ref{fig:field_compare}
with the field sample from \citetalias{sco16}, which derived the average gas properties of galaxies stacked in bins of stellar mass and SSFR.  For a fair comparison, we limit our comparison to their stacks at $z\sim1$ in the range log $M_{\star}/\Msun> 10.5$ and SFR$\,>50\,\Msun$ yr$^{-2}$, shown as blue diamonds. The average gas masses of our cluster stacks (red cross and gold x) fall well below the general distribution of the gas masses of the field galaxies in the \citetalias{sco16} sample.  When incorporating SFR and stellar mass in the gas depletion timescales and gas fractions, cluster galaxies are, on average, consistent with the shortest gas depletion timescales and lowest gas fractions as found at $z\sim1$ for massive field SFGs.

To quantify this difference, we rebin the \citetalias{sco16} stacks into two bins split by the median stellar mass of our sample (Table~\ref{tbl:properties}).  For these bins, we determine the weighted average and standard deviation of the \citetalias{sco16} stacks, which each include 1-12 galaxies, adopting as weights the number of galaxies in each stack.  The rebinned average gas properties can be seen in the first column of Figure~\ref{fig:field_compare} (indicated by the blue background).  To calculate the deficits, we then take the ratio of the rebinned field to our cluster stacks, finding that the clusters are lower than the field in $M_{\rm mol}$ by $2.7\pm1.0$ ($2.3\pm0.9$), in $\tau_{\rm depl}$ by $3.8\pm1.6$ ($2.9\pm0.7$), and in f$_{\rm gas}$ by $2.0\pm0.6$ ($3.4\pm0.8$) in the low (high) mass bin (Table~\ref{tbl:deficits}).

We next expand our comparison to the field scaling relations, which are widely used in the literature and allow us to compare to a predicted field value calculated for the average properties (stellar mass, etc) of our stacked subsamples.  We adopt the scaling relations presented in \citet{sco17} (hereafter S17), which were derived from a sample of Herschel-selected SFGs at a range of redshifts ($z\lesssim3$) and masses (log $M_{\star}/\Msun>10.3$).  Priors were used to push the extraction of the Herschel photometry to lower significance detections in a similar manner as used in our Herschel photometry \citepalias{alb16}. Both the \citetalias{sco17} field and our cluster samples are the massive, IR-bright end of the galaxy populations, though the use of detections, not stacking, in the field scaling relations may bias the results toward the gas-rich populations.  We note that for the stellar mass range probed in this work, the \citetalias{sco17} scaling relations are comparable to those of \citet{tac18, tac20}, which combine CO and dust continuum emission measurements, both detected and stacked.  

In Figure~\ref{fig:field_compare} (all panels), we show the field scaling relations for $M_{\rm mol}$, $\tau_{\rm depl}$, or $f_{\rm gas}$ as a function of stellar mass, redshift, SFR$_{\rm IR}$, or $\Delta$MS $-$ with the other parameters being fixed at the average values of our stacked subsets (Table~\ref{tbl:properties}) $-$ following Eqns 6-8, 10 in \citetalias{sco17}.  For example, in the upper lefthand plot, we show the gas mass of a field galaxy as a function of stellar mass, with its redshift and distance from the MS fixed to the properties of our lower stellar mass stack.  We repeat this for the higher mass stack.
The deficit for each cluster stack is then calculated as the ratio between the predicted field value and the stacked value.  The error on the deficit is a combination of the error on the best-fit scaling relations\footnote{The error on the S17 scaling relations shown is derived using the errors on the parameters defining the best-fit relations. This does not include all possible calibration and systematic errors, which are expected to be largely shared between the cluster and field gas mass derivations and so are not included in the comparison. } and the bootstrapped errors of the stacks, which included the spread in the population, added in quadrature.

From this analysis, we can compare the slope of our cluster stacks (split into two bins by stellar mass, etc) against the slope of the field scaling relations as well as the absolute values.
In terms of gas mass (top row), the cluster galaxies show similar trends as the field in terms of increasing gas mass with redshift, stellar mass, and obscured SFR, but at a deficit of 3-5x in absolute value.  Cluster galaxies with $\Delta$MS$\,>0.35$ show a hint of a reverse trend in gas mass from the field scaling relation, with a larger deficit for this population relative to cluster galaxies closer to $\Delta$MS$\,\sim0$. This is likely driven by the fact that our starbursting galaxies ($\Delta\textrm{MS}>0.6$) are predominantly the lowest mass galaxies in our sample.  This is a selection effect caused by targeting sources based on Herschel flux; at the low mass end, only starbursts will be bright enough to make it into the sample. This is confirmed when we look at the gas fraction as a function of $\Delta$MS (fourth column, third row), which looks at the gas content for a given stellar mass.  We find that the slope of the field scaling relation with $\Delta$MS is recovered, though the absolute deficit in cluster gas remains.  This suggests that there is no preferential loss of gas in currently starbursting cluster galaxies, though we also note that high mass, extremely dusty starbursts may still be missing from our sample due to the difficulty in measuring a robust photo-$z$.  CO spectroscopy is needed to place the extremely dusty sources in the cluster and measure their gas properties.

The gas depletion timescales (second row) and gas fractions (third row) are similarly consistent with the slopes expected in the field when splitting by the properties shown.  The deficits are significant, however, with average gas depletion timescales of $\sim200$ Myr, $\sim5-7$x lower than the predicted field values, and gas fractions of 10-20$\%$, a deficit of $\sim2-3$x.  All stacked values are listed in Table~\ref{tbl:properties} and the deficit values are in Table~\ref{tbl:deficits}.

\subsection{Comparison with other (proto-)clusters: pushing past the gas-rich outliers}\label{sec:cluster_compare}

In this section, we compare to the gas properties in cluster galaxies in the literature derived from both CO and dust continuum, which are both robust tracers of the gas mass \citep[][]{tac20}.  We do not re-calibrate any measurements based on CO and we caution that, similar to the comparison to field galaxies, comparisons between cluster samples can suffer from heterogeneous datasets and methodologies, as well as biased selection via targeted follow-up, small number statistics, and often high detection limits \citep[e.g. log $M_{\rm mol, CO}/\Msun>10.8$ at $z\sim1.5$;][]{nob17, hay17, hay18, sta17}. One advantage we have in this work is that we draw targets from eleven clusters with a range of star formation activity \citepalias{alb16}, mitigating the chances that we are biased toward only highly star forming cluster systems.  Of course, on the other hand, our target selection of luminous cluster SFGs  introduces a different bias and means we can only address the gas properties of that population.

\begin{figure}
    \centering
    \includegraphics[width=\linewidth, trim=5 0 5 0, clip]{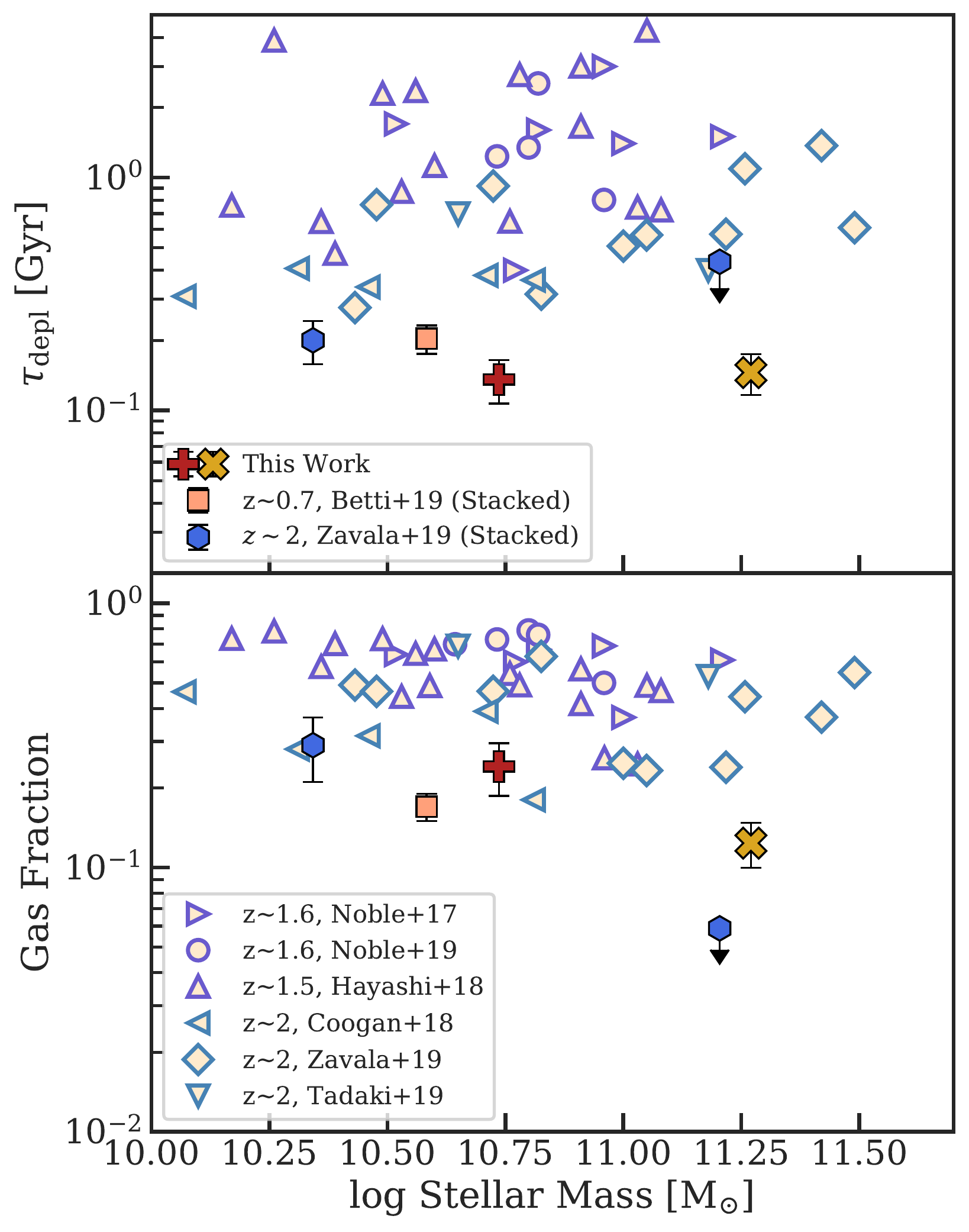}
    \caption{The gas depletion timescale (top) and gas fraction (bottom) as a function of stellar mass for (proto-)cluster galaxies.  The red plus and yellow x are the stacked results from this work, split by the median stellar mass of the sample (Table~\ref{tbl:properties}). We compare to the gas properties of cluster galaxies stacked in dust continuum at $z\sim0.7$ \citep[orange square,][]{bet19} and $z\sim2$ \citep[light blue hexagons,][]{zav19}.  Individual detections of CO and/or dust emission in (proto-)cluster galaxies are shown at $z\sim1.5$ \citep[purple circles and upright and rightward triangles;][]{nob17, nob19, hay17, hay18} and $z\sim2$ \citep[blue left triangles, diamonds, and inverted triangles;][]{coo18, zav19, tad19}.  The stacked (average) values of cluster galaxies indicate that the typical gas content is lower than implied by detections, with shorter depletion timescales and lower gas fractions.}
    \label{fig:cluster_comparison}
\end{figure}

Closely analogous to this work in terms of technique, \citet{bet19} looked at the stacked dust continuum emission in 101 SFGs at $z\sim0.7$ as a function of local galaxy density \citep{sco13}, with their highest density bin corresponding to the cluster environment.  In this lower redshift epoch ($z<1$), the quenching efficiency in clusters is high \citep{muz12, alb14, nan17}; however, among cluster galaxies \textit{with ongoing star formation}, the average gas properties look remarkably similar at $z\sim0.7$ and $z\sim1.4$ (this work), with short depletion timescales and low gas fractions (Figure~\ref{fig:cluster_comparison}).

At more comparable redshifts ($z\sim1.5$), relatively small samples of cluster galaxies have been studied via detections of the CO emission line or dust continuum emission.  \citet{nob17} looked at 11 cluster members detected in CO(2-1) in three clusters at $z\sim1.6$, finding field-like gas depletion timescales and both field-like and {\it enhanced} gas fractions.  \citet{hay17, hay18} looked at CO(2-1) and dust continuum for 18 cluster galaxies, again finding enhanced gas fractions and long depletion timescales compared to field scaling relations.  They speculated that gas accretion may be enhanced in cluster infall regions and/or filaments or that SFEs may be reduced due to environmentally-induced shock-heating or feedback.  These studies, however, have relatively high detection limits and may be missing the gas-poor cluster population. 

Interestingly, however, we have to date two examples of detection studies at $z\sim1.5$ which have gone significantly deeper, albeit over small areas.  \citet{nob19} presented deep follow-up of one of their \citet{nob17} clusters, detecting four additional cluster galaxies in CO(2-1) (Figure~\ref{fig:cluster_comparison}).  Additionally, a serendipitous detection of CO in a low mass cluster at $z\sim1.3$ was recently obtained to a $5\sigma$ limit of log $M_{\rm mol}/\Msun\sim10.2$; Williams, et al., ApJ, submitted).  Despite probing deeper in terms of $M_{\rm mol}$, neither of these studies uncovered the relatively gas-poor cluster galaxies suggested by our stacking analysis.  In fact, both studies found only cluster galaxies with gas comparable to the upper end or above the field scaling relations presented in \citet{tac18}, which are built largely on CO detections. 

A cluster study at higher redshift, on the other hand, does finds a gas-poor population:  \citet{coo18} obtained deep CO(1-0) from the VLA and dust continuum emission at 870$\mu$m in the core of a low mass, X-ray-selected cluster at $z=1.99$, which revealed relatively low gas fractions, some comparable to this work, in five galaxies.  Gas depletion timescales and gas fractions derived from their dust continuum imaging (which are in good agreement with their CO (1-0) detections and limits) are shown in Figure~\ref{fig:cluster_comparison}.  Notably, their galaxies are on the MS in terms of their SSFRs, but have high gas excitation (obtained from multiple CO transitions) and a high rate of mergers/interactions and/or AGN activity.

At higher redshifts ($z>2$), gas content has been studied in the cores of overdense proto-cluster environments \citep[i.e.][]{wan16, wan18, ume17, zav19, tad19, cha21, lon20}.  We limit our comparison to two studies at $z\sim2$, \citet{tad19} and \citet{zav19}, with the caveat that these environments may be in a different stage of virialization.  \citet{tad19} looked at 13 H$\alpha$-selected proto-cluster members in CO(3-2) around the radio galaxy PKS 1138−262 at $z=2.16$, again finding enhanced gas fractions and long depletion timescales in four detections.  \citet{zav19} looked at 41 proto-cluster SFGs with an average mass of log $M_{\star}/\Msun = 10.8$ at $z\sim2.1$ in a log $M_{\rm halo}/\Msun=14.3$ proto-cluster core, finding again similar gas content as coeval field galaxies, but with increasing gas-poor cluster members at the high mass (log $M_{\star}/\Msun>11$) end.  \citet{zav19} additionally performed stacking to better characterize their non-detections, finding average values on the low end of their detected distribution, consistent with this work and \citet{bet19}.  This suggests that the (massive) cores of proto-clusters at $z\sim2$ may also be experiencing environmental gas loss.

Putting this all together in Figure~\ref{fig:cluster_comparison}, and folding in the field comparisons of Section~\ref{sec:field_compare},  it is clear that the picture of field-like gas content from cluster galaxy detections, with long depletion timescales and enhanced gas fractions, does not describe star forming cluster galaxies on average.  Instead these gas-rich populations may be outliers driven by high detection limits.  On the other hand, the two deep cluster observations at $z\sim1.5$ that have been obtained to date \citep[][Williams et al, ApJ, submitted]{nob19} have also found gas-rich cluster populations, deepening the mystery.  We have only one example of relative gas-poor detections at $z=1.99$ \citep{coo18}.  Resolving this contention between detections and stacking will require even lower detection limits to recover the distribution of gas properties in the galaxies represented in the stacks. Care must additionally be taken to avoid biases by observing different regions (i.e. core, infalling) within the cluster environment as well as statistical clusters samples, given cluster-to-cluster variation.


\section{Discussion}\label{sec:disc}

\subsection{Warm Dust in Cluster Galaxies}\label{sec:warm_dust}

Dust temperature has been linked to internal factors such as the physical conditions of SF in a galaxy \citep{gal18}; however, its dependence on external factors such as environment remain poorly constrained.  In Section~\ref{sec:ir_sed}, we measured the average dust temperature of our cluster galaxies via their Herschel+ALMA SEDs modeled with a modified blackbody and power law component \citep{cas12}.  We found that for all three radial bins, out to twice the virial radius, the dust temperatures are relatively warm, $\sim36-38$ K.  Comparably massive field galaxies have typical dust temperatures of $30\pm4$K with relatively little scatter \citep[][see also \citet{sym13}]{sch18}.  We note this comparison assumes that the dust emissivity $\beta$ is comparable in cluster and field galaxies, something that has yet to be tested in the literature.  Similarly warm dust temperatures have been observed in a small fraction ($\sim10\%$) of sub-LIRG cluster members in the Bullet cluster,  in excess of what is expected in the field and attributed to dust stripping and heating from RPS \citep{raw12}.   FIR stacking of a much larger sample selected from SDSS found a marginal trend of increasing dust temperature with increasing environment, up to filament and cluster scales \citep{mat17}.  Our results qualitatively support this trend at higher redshift, with the caveat that the conditions that set the dust temperature also evolve with redshift and stellar mass \citep{sch18} so the absolute numbers cannot be easily compared.  Conversely, studies in both the Coma cluster \citep{ful16} and at higher redshift \citep{nob16} found no increase in dust temperature in cluster environments; however, long wavelength constraints were often undetected or not included in these works.  A systematic study with good FIR and submm wavelength coverage is needed.  The inclusion of the ALMA datapoint was key in our analysis; without it, we had previously found a good match to the Herschel data only using templates with $T_{\rm dust}\lesssim30$ K for the cold dust component \citep{kir15}.

\subsection{Cold Gas in Cluster Galaxies: Observations and Theory}\label{sec:theory}

In this work, we have quantified the gas properties of a population of massive cluster SFGs at $z\sim1.4$ which are on or above the star forming MS.  We have found a clear dependence of their gas properties on the environment, with gas masses 2-3x (3-5x) lower than stacked co-eval field galaxies (field scaling relations), with correspondingly short gas depletion timescales and low gas fractions.  This effect is seen out to twice the virial radius, the limit of our survey.

Significant gas loss starting at large cluster-centric radii has been predicted by recent cosmological simulations. \citet{zin18} studied the effects of RPS on diffuse hot halo gas and the more tightly bound cold disk gas.  They found that RPS can efficiently remove 40-70$\%$ of the galaxy's hot halo gas at large cluster-centric radii ($\sim2R_{\rm vir}$), with less than 30$\%$ of halo gas remaining in high mass satellites by $R_{\rm vir}$ \citep[see also][]{bah13}. This removal was attributed to virial accretion shocks \citep{sar98, bir03, dek06} at roughly the boundary where galaxies enter the hot ICM.  This is also approximately where it is expected that fresh gas accretion onto galaxies is prevented (i.e. starvation).  For their high redshift clusters ($z\sim0.6$), \citet{zin18} estimated a travel time of a few Gyr between the accretion shock and $R_{\rm vir}$ over which this starvation can occur. If we compare this to the expected gas recycling timescale \citep[$0.5-1\,$Gyr;][]{opp10, tac20} and gas depletion timescale due to ongoing SF  \citep[$\sim1$ Gyr at $z\sim0.6$;][]{tac20} for field galaxies, then there is ample time for the combination of RPS of the hot halo gas, starvation, and consumption of gas in star formation to affect the cold gas reservoir well before the galaxy reaches the virial radius.
This is qualitatively consistent with our recovery of a gas deficit out to 2$R_{\rm vir}$.  Additionally, cold gas may be heated by processes such as feedback, potentially associated with increase fractions of cluster AGN \citep[][A16]{mar13}.  

Similar but more extreme results were found in the \textsc{Three Hundred} simulation suite \citep{art19, mos21}, which looked at cluster and satellite halos up to $z\sim1$.  They predict instantaneous gas fractions consistent with zero by the crossing of $R_{\rm vir}$ in projected space, again attributed to crossing virial accretion shocks.  This total gas loss is likely overestimated due to poor resolution in the current simulations \citep{bah17b, bos18b, bos18a} and oversimplifications in ICM structure \citep{ton19}, but the emerging picture is that of significant gas loss starting at large radii and largely completing by the first passage of the cluster center, largely independent of halo mass \citep[see also][]{oma16, lot19, oma21}.  

Our finding of gas properties 2-5x below the field, on average, out to $2R_{\rm vir}$ supports this picture. The low gas content can be reconciled with the ongoing star formation under the ``delayed, then rapid" quenching scenario \citep[i.e.][]{wet13, vdb13, bah15, oma16, alb16, nan17, rhe20} in which any effect on the SFR is delayed after infall, followed by quenching on a short timescale. If tightly bound disk gas is retained, then the high mass galaxies characteristic of this study may then keep forming stars up to a Gyr after infall \citep{lot19}. The low gas masses, ongoing star formation, and condition that quenching is rapid, particularly for these clusters during the era in which we see a sharp rise in the quenched fraction over $z\sim1-1.6$ \citep[A16,][]{nan17}, suggest a combination of gas loss via stripping to prevent the recycling and/or cooling of gas  \citep[e.g. effected by stellar or AGN winds;][]{tac20} in conjunction with the consumption of disk gas via star formation, facilitates the final quenching.


Seemingly at odds with the low average gas content presented in this study $-$ and the simulations that predict strong gas stripping on first passage $-$ are the gas-rich cluster galaxies discussed in Section~\ref{sec:cluster_compare}.  These galaxies must 1) retain (or replenish) their cold gas reserves or 2) be in a phase of efficient gas cooling, a selection effect. For the former, hot gas haloes have been observed around local cluster/group galaxies \citep{sun07, jel08}.   One explanation is that these individually detected, gas-rich cluster galaxies occupy a privileged place in the clusters relative to gas streams which may weave through the ICM and could provide fresh gas accretion \citep{dek06, zin16, zin18}. Such gas streams, particularly in unrelaxed clusters, have been invoked to explain metal enrichment of the ICM at large radii \citep{reb06,sim15} and the build up of Brightest Cluster Galaxies \citep[BCGs;][]{mcd12}.  This work quantifying the average gas properties of massive cluster SFGs, drawn from a range of clusters, adds a key constraint for future investigations of this mechanism, in that we have shown that this gas-rich population is the minority.

\section{Conclusions} \label{sec:conc}

In this work, we have used ALMA Band 6 (observed 1.3mm) imaging to present the average gas properties of 126 massive (log $M_{\star}/\Msun\gtrsim10.5$) cluster SFGs in eleven massive (log $M_{\rm halo}/\Msun\sim14$) galaxy clusters at $z=1-1.75$.  Our eleven clusters represent a range in total star formation activity and we sample out to $\sim2 R_{\rm vir}$, with $R_{\rm vir}$ likely equivalent to the splashback radius. On initial analysis, it was found that all targets were undetected in ALMA due to weaker submm emission relative to that predicted based on coeval field galaxies.  Stacking analysis was therefore used to obtain the average dust continuum emission, from which we derived average gas masses ($M_{\rm mol}$) for stacked subsets of our sample following the calibrations presented in \citetalias{sco16}.  Combined with extensive multi-wavelength data, we further measured the average gas depletion timescales ($\tau_{\rm depl}$; inversely, the star formation efficiencies) and gas fractions (f$_{\rm gas}$).

Our main conclusions are as follows:

\begin{enumerate}
    \item ALMA stacks in three cluster-centric radial bins probing from the virialized to infalling regions were combined with the Herschel/PACS photometry to construct the average IR SEDs of massive, star forming cluster galaxies.  We find weaker submm flux than predicted by the IR SEDs of ALMA-detected field galaxies at $z\sim2$ \citep{kir15, dun17}; our cluster SEDs are instead well described by local SFG templates \citep{rie09} of similar luminosity (log $L_{\rm IR}/\Lsun\sim11.9$). The addition of the submm anchor to the Herschel data at the dust peak reveals relatively warm dust temperatures ($\sim36-38$ K) compared to the well constrained cooler temperatures of field galaxies \citep[$\sim30$ K;][]{sch18}, with no significant dependence on cluster-centric radius.  Evidence of warmer dust in cluster galaxies is highly conflicted in the literature; in our work, it was key to probe both the IR peak and submm to identify these warmer temperatures.
    
    \item Gas masses, gas depletion timescales, and gas fractions were presented in the three radial bins, sampling out to twice the virial radius.  We find that the gas properties have weak to no dependence on cluster-centric radius, with a marginal excess in average gas mass in the intermediate bin ($\sim R_{\rm vir}$) at the $1\sigma$ level.  We note that using projected radial bins may dilute a real trend in this analysis; however, it is clear that the environmental impact on gas properties extends from the cluster cores to large radii, in good agreement with cluster studies which have traced the cluster influence beyond the virial (splashback) radius \citep[i.e][]{bal99, vdl10, chu11, ras12, shi21}.
    \item Restacking our cluster sample into bins split by stellar mass, redshift, obscured SFR, and distance from the MS, we perform a careful comparison of our ALMA cluster stacks to the gas properties of field galaxies in the literature.  We find that our cluster galaxies have deficits of 2-3x, 3-4x, and 2-4x in $M_{\rm mol}$, $\tau_{\rm depl}$, and f$_{\rm gas}$ when compared to stacked field samples \citepalias{sco16}.  Slightly larger deficits are found when comparing to the predicted values from field scaling relations \citep{tac18, sco17}, calculated based on the median properties (redshift, stellar mass, etc) of our subsamples.  This demonstrates that environmentally-driven processes are causing, on average, a significant depletion in the gas reservoirs of cluster galaxies at high redshift.
    
    \item Comparison of our stacks with gas measurements of cluster galaxies in the literature reveal that our results are consistent with other stacking analyses \citep{bet19, zav19}, which also find on average low gas content.  By contrast, our results are inconsistent with studies of cluster populations {\it detected} in CO or dust continuum emission, which largely find field-like gas depletion timescales and field-like or even enhanced gas fractions \citep[i.e.][Williams et al., ApJ, submitted; but see \citet{coo18}]{hay17, hay18, nob17, nob19, tad19, zav19}. 
    The comparison with our average, stacked gas properties suggests that these gas-rich cluster galaxies may not be typical.
    
\end{enumerate}

Our observed deficit in the gas properties of cluster galaxies at $z\sim1.4$ out to $2R_{\rm vir}$ is qualitatively in good agreement with recent simulations, which project that gas depletion begins early in the infall stages, at large radii where galaxies encounter virial accretion shocks.  Though they vary in detail and quantity of gas removed, these simulations largely agree that significant gas stripping happens by first passage of the cluster center \citep{zin18, art19, mos21, oma21}.  Our analysis of star-forming yet gas-poor cluster galaxies is consistent with this significant gas loss, but also a retention of the closely-bound disk gas, which continues for a time to fuel star formation as in the ``delayed, then rapid" quenching scenario \citep[i.e.][]{wet13}.

\acknowledgments

The authors thank Sarah Betti for her help in measuring the ALMA photometry and Karen Olsen for discussions on cluster simulations.  SA acknowledges support from the James Webb Space Telescope (JWST) Mid-Infrared Instrument (MIRI) Science Team Lead, Grant 80NSSC18K0555, from NASA Goddard Space Flight Center to the University of Arizona.  CCW  acknowledges  support from the JWST Near-Infrared Camera (NIRCam) Development  Contract  NAS5-02105  from  NASA  Goddard  Space  Flight Center to the University of Arizona.  The National Radio Astronomy Observatory is a facility of the National Science
Foundation operated under cooperative agreement by Associated Universities, Inc. This paper makes use of the following ALMA data: ADS/JAO.ALMA\#2015.1.00813.S. ALMA is a partnership of ESO (representing its member states), NSF (USA) and NINS (Japan), together with NRC (Canada), NSC and ASIAA (Taiwan), and KASI (Republic of Korea), in cooperation with the Republic of Chile. The Joint ALMA Observatory is operated by ESO, AUI/NRAO and NAOJ.  This research was carried out in part at the Jet Propulsion Laboratory, California Institute of Technology, under a contract with the National Aeronautics and Space Administration.
\\
\\
Software: NumPy \citep{har20},
Matplotlib \citep{hun07}, Astropy \citep{astropy:2018}, pandas \citep{mckinney-proc-scipy-2010}, seaborn \citep{Waskom2021}, CASA \citep{mcm07}

\bibliographystyle{aasjournal} 
\bibliography{main} 



\end{document}